\documentclass[apj]{emulateapj}

\begin{document}

\title{On the inconsistency between the black hole mass function \\ 
inferred from $M_\bullet-\sigma$ and $M_\bullet-L$ correlations}

\author{Elena Tundo\altaffilmark{1,2}, 
        Mariangela Bernardi\altaffilmark{1}, 
        Joseph B. Hyde\altaffilmark{1},
        Ravi K. Sheth\altaffilmark{1}, \\
        and Alessandro Pizzella\altaffilmark{2}}

\affil{}

\altaffiltext{1}{Dept. of Physics and Astronomy, University of Pennsylvania, 
                 209 South 33rd St, Philadelphia, PA 19104, U.S.A.}

\altaffiltext{2}{Dipartimento di Astronomia, Universita' di Padova, 
                 vicolo dell'Osservatorio 3/2 I-35122, Padova, Italy}

\begin{abstract}
 Black hole masses are tightly correlated with the stellar 
 velocity dispersions of the bulges which surround them, and 
 slightly less-well correlated with the bulge luminosity. 
 It is common to use these correlations to estimate the expected 
 abundance of massive black holes.  This is usually done by 
 starting from an observed distribution of velocity dispersions 
 or luminosities and then changing variables.  
 This procedure neglects the fact that there is intrinsic scatter 
 in these black hole mass--observable correlations.  Accounting 
 for this scatter results in estimates of black hole abundances 
 which are larger by almost an order of magnitude at masses 
 $>10^{9}M_\odot$. Including this scatter is particularly 
 important for models which seek to infer quasar lifetimes and 
 duty cycles from the local black hole mass function. However, 
 even when scatter has been accounted for, the $M_\bullet-\sigma$ 
 relation predicts fewer massive black holes than does the 
 $M_\bullet-L$ relation.  This is because the $\sigma-L$ relation 
 in the black hole samples currently available is inconsistent with 
 that in the SDSS sample from which the distributions of $L$ or 
 $\sigma$ are based:  the black hole samples have smaller $L$ for 
 a given $\sigma$, or larger $\sigma$ for a given $L$.  
 The $\sigma-L$ relation in the black hole samples is similarly 
 discrepant with that in other samples of nearby early-type 
 galaxies.  This suggests that current black hole samples are biased:  
 if this is a selection rather  than physical effect, then the 
 $M_\bullet-\sigma$ and $M_\bullet-L$  relations currently in the 
 literature are also biased from their true values.  
\end{abstract}

\keywords{galaxies: elliptical --- galaxies: fundamental parameters --- 
black hole physics}

\section{Introduction}
The abundance of supermassive black holes is the subject of 
considerable current interest 
(e.g.  Yu \& Tremaine 2002; Marconi et al. 2004; 
McLure \& Dunlop 2004; Shankar et al. 2004; Yu \& Lu 2004; 
Ferrarese \& Ford 2005).  
Several groups have noted that galaxy formation and supermassive black 
holes growth should be linked, and many have modeled the joint 
cosmological evolution of quasars and galaxies (see,
e.g., Monaco et al. 2000; Kauffmann \& Haehnelt 2001; Granato et al. 2001; 
Cavaliere \& Vittorini 2002; Cattaneo \& Bernardi 2003; Haiman et al. 2004;
Hopkins et al. 2006; Lapi et al. 2006; Haiman et al. 2006 and references 
therein).  
Since the number of black hole detections to date is less than 
fifty, their abundance is estimated by using secondary indicators.  
In particular, $M_\bullet$ is observed to correlate strongly and 
tightly with the velocity dispersion of the surrounding bulge 
(e.g. Ferrarese \& Merritt 2000; Gebhardt et al. 2000; Tremaine et al. 2002).  
Since detecting bulges is considerably easier than detecting black 
holes, it has become common to estimate the abundance of black holes 
by combining the observed distribution of bulge velocity dispersions 
(e.g. Sheth et al. 2003) with the observed $M_\bullet-\sigma$ 
relation.  A crude estimate follows easily if one is willing to 
assume that all bulges host black holes, and that the 
$M_\bullet-\sigma$ relation has no intrinsic scatter 
(e.g. Yu \& Tremaine 2002; Aller \& Richstone 2002).

There is some discussion in the literature about whether $L$ or 
$\sigma$ is a better predictor of $M_\bullet$.  There are two 
parts to this statement which are not always stated explicitly.  
The first is the assumption that the $M_\bullet-$observable 
relation is a single power law; whether this is a better 
approximation for $L$ than for $\sigma$ is an open question, 
although Lauer et al. (2007) argue that the curvature in the 
$\sigma-L$ relation for massive early-type galaxies 
(Bernardi et al. 2007a; Bernardi 2007) suggests 
that the $M_\bullet-\sigma$ relation is unlikely to be a 
single power-law.  
In what follows, we will assume the relations in question are 
indeed single power laws.  

The second is the issue of the scatter around the mean relations.  
It is generally believed that the relation with smaller scatter 
provides the better estimate of the $M_\bullet$ distribution.  
Indeed, Marconi et al. (2004) state that if the scatter around 
two relations is similar, then both relations should provide 
equivalent descriptions of the distribution of $M_\bullet$.  
One of the goals of the present paper is to show that this is not 
the whole story.  Provided the intrinsic scatter around the two 
relations is accurately known, whether or not one relation is tighter 
than another is irrelevant.  (The only practical difference is that, 
if the intrinsic scatter is smaller, then observations of fewer objects 
are required to estimate it reliably.)  

Both the $M_\bullet-\sigma$ and $M_\bullet-L$ relations show 
considerable scatter, not all of which can be accounted-for by 
measurement errors.  Marconi \& Hunt (2003) present evidence that 
the amount by which an object scatters from these relations is 
correlated with bulge size (half light radius), suggesting that 
at least some component of the scatter is intrinsic.  
Gebhardt et al. (2000) suggest that the intrinsic scatter in 
$M_\bullet$ at fixed velocity dispersion is of order 0.25~dex, 
whereas scatter around $M_\bullet-L_V$ is about 0.35~dex 
(e.g. Novak et al. 2006).  
If the intrinsic scatter is indeed this large, then it must be 
accounted for, especially when estimating the abundances of the 
most massive black holes ($M_\bullet \ge 10^9\,M_\odot$).  

Section~2 describes a toy model of the effects of scatter which 
shows that,
 (i) if intrinsic scatter is ignored, then both the $L$- and 
  $\sigma$-based predictions will underestimate the true abundance  
  of the most massive black holes; 
 (ii) the observable which correlates most tightly with $M_\bullet$ 
  will provide the best estimate of the true abundance of the most 
  massive black holes; 
 (iii) if scatter has been correctly accounted for, $\sigma$- and 
  $L$-based predictors of $M_\bullet$ abundances should give the 
  same answer.  
It then shows the $M_\bullet-\sigma$, $M_\bullet-L$ and $\sigma-L$ 
correlations, their scatter, and how we use them to estimate 
black hole abundances.  A direct comparison of the luminosity and 
velocity dispersion based predictors is provided, both when  
intrinsic scatter in these relations is accounted for and when it 
is ignored.  

We find that, if scatter is ignored, then the $L$-based method 
predicts substantially more $10^9 M_\odot$ objects than does the 
$\sigma$-based method.  The toy model suggests that this may be 
a consequence of ignoring the intrinsic scatter.  However, accounting 
for this scatter does not eliminate this discrepancy, suggesting that 
there may be a more serious inconsistency.  
Section~3 identifies the reason for this discrepancy with the fact 
that the $\sigma-L$ relation in the SDSS/Bernardi et al. (2003a)
(hereafter SDSS-B07; see Bernardi 2007 for the definition of the SDSS-B07 
sample), from which the $L$ and $\sigma$ distributions are drawn, is 
rather different from that in the black hole samples, from which the 
$M_\bullet-\sigma$ and $M_\bullet-L$ relations are derived.  
A more detailed analysis of the role of selection effects in the 
$M_\bullet$ sample is presented in Bernardi et al. (2007b). 
A final section discusses our findings and summarizes our 
conclusions.  
A standard flat $\Lambda$CDM comological model has been used, 
with $\Omega_m=0.3$, $\Omega_\Lambda=0.7$ and 
$H_0=70$~km~s$^{-1}$~Mpc$^{-1}$.  

\section{Black hole abundances from $M_\bullet$-observable correlations}
The first part of this section discusses the effect of intrinsic 
scatter in $M_\bullet$-observable relations on inferences about 
black hole abundances.  The second part shows various 
$M_\bullet$-observable correlations in the compilation of 
H\"aring \& Rix (2004).  
The third and fourth parts of this section show the predicted black 
hole abundances when intrinsic scatter in these relations is 
accounted for and when it is not.

A detailed discussion of exactly how the black hole sample was 
compiled, as well as how we convert from B, V, R and I-band 
luminosities to SDSS $r-$band is provided in Appendix~A of 
Bernardi et al. (2007b).  Briefly, all luminosities and black 
hole mass estimates depend on distance:  where necessary, 
these were computed by scaling results in the literature to 
$H_0 = 70$~km~s$^{-1}$~Mpc$^{-1}$.  
The estimated velocity dispersions are, essentially, distance 
independent (see Bernardi et al. 2007b for details).

\subsection{A simple model of the effect of intrinsic scatter}
Consider three observables which we will call $L$, $V$ and $M_\bullet$, 
with joint distribution $p(L,V,M_\bullet)$.  To make the discussion 
more concrete, suppose that this joint distribution is Gaussian, so 
that this distribution is completely specified by the means and 
variances of the three variables, and the three cross-correlation 
coefficients $r_{\rm VM_\bullet}$, $r_{\rm LM_\bullet}$, and 
$r_{\rm LV}$.  These correlation coefficients are constrained to 
lie between $\pm 1$, with a value of zero indicating no correlation.  
Then the distribution of $M_\bullet$ at fixed $O$, with $O=L$ or $V$, 
is Gaussian with mean and variance 
\begin{eqnarray}
 \langle M_\bullet|O\rangle 
  &=& \langle M_\bullet\rangle + 
   r_{OM_\bullet}\,\sigma_{M_\bullet}\,(O-\langle O\rangle)/\sigma_O, 
 \label{mean}\\
 \sigma^2_{M_\bullet|O} &=& \sigma^2_{M_\bullet}\,(1-r_{OM_\bullet}^2).
 \label{var}
\end{eqnarray}
Let $p_O(M_\bullet)$ denote the result of predicting the 
distribution of $M_\bullet$ from the distribution of $O$ by using 
$\langle M_\bullet|O\rangle$ to change variables from 
$p(O)dO = p_O(M_\bullet)dM_\bullet$.  Then 
$p_O(M_\bullet)$ is a Gaussian centered on $\langle M_\bullet\rangle$ 
with rms = $|r_{OM_\bullet}|\sigma_{M_\bullet}$.  
Unless $r_{OM_\bullet}=\pm 1$, this value will be smaller than 
$\sigma_{M_\bullet}$.  Thus, in general, 
 (i) $p_V(M_\bullet)\ne p_L(M_\bullet)$ and 
 (ii) both will be more sharply peaked than the true $p(M_\bullet)$ 
distribution.  (In the limit $r_{OM_\bullet}\to 0$, i.e., the limit 
of no correlation between O and $M_\bullet$, $p_O(M_\bullet)$ 
becomes a delta function centered on the mean value; this behaviour 
is the basis for the concept of `shrinkage towards the mean' which 
is common in discussions of Bayesian statitical inference. )  
Hence, except in the case of perfect correlation between $M_\bullet$ 
and $O$, all choices of $O$ are biased---there is little reason to 
prefer the estimate from one observable over another.  

On the other hand, although both $p_V(M_\bullet)$ and $p_L(M_\bullet)$ 
will underestimate the true distribution $p(M_\bullet)$ at large 
$M_\bullet$, the discussion above shows that the distribution of 
the observable which correlates more strongly with $M_\bullet$ will 
be closer to the true $p(M_\bullet)$.  In particular, at large 
$M_\bullet$, the cumulative distribution of the observable which 
correlates more strongly with $M_\bullet$ will be closer to the 
true $p(>M_\bullet)$.  So one might argue that 
the observable which predicts the largest $p_O(M_\bullet)$ at the 
largest $M_\bullet$ is the one which is closest to yielding the 
true value.  
(Of course, this is only true in an ideal world in which there 
are no systematic measurement errors.)  

In effect, the procedure just described ignores the scatter around 
the mean $\langle M_\bullet|O\rangle$ relation.  
To include the effects of this scatter one must convolve $\phi(O)$ 
with the distribution $p(M_\bullet|O)$ which has mean 
$\langle M_\bullet|O\rangle$ and rms $\sigma_{M_\bullet|O}$:  
\begin{equation}
 \phi(M_\bullet) \equiv \int dO\,\phi(O)\,p(M_\bullet|O) 
 \label{scat}
\end{equation}
Provided $\langle M_\bullet|O\rangle$ and $\sigma_{M_\bullet|O}$ 
are accurately known, it doesn't matter what $O$ is, or how 
tightly correlated it is with $M_\bullet$.  That is to say, 
predicting the distribution of $M_\bullet$ from $L$ using the 
expression above should give the same (correct) answer as predicting 
it from $V$.  

If this does {\em not} happen, i.e., if the setting of $O=L$ 
gives a different answer than $O=V$, then this is an indication 
that one or more of the $p(M_\bullet|O)$ relations are incorrect.  
This may happen, for instance, if $\phi(L)$ and $\phi(V)$ are 
estimated from a different dataset from which the $M_\bullet-L$ 
and $M_\bullet-V$ correlations are estimated, since, if the datasets 
are not the same, then there is no guarantee that the joint 
$M_\bullet-L-V$ distributions in the two datasets are the same.  
We argue below (see Section~3) that this appears to be the case:  the $V-L$ 
correlation defined by the black hole samples in the literature 
differs from that in the SDSS-B07, which currently offers the best 
determinations of $\phi(L)$, $\phi(V)$ and perhaps also $V-L$ 
(see Bernardi et al. 2007b).  

\subsection{The $\langle M_\bullet|L\rangle$ and 
                $\langle M_\bullet|\sigma\rangle$ relations}
The discussion above makes clear that, if $O$ is to predict $M_\bullet$, 
then the correlation of interest is $\langle M_\bullet|O\rangle$.  
Use of the (inverse of the slope of the) $\langle O|M_\bullet\rangle$ 
correlation for this purpose is clearly incorrect.  For similar 
reasons, it is logically inconsistent to use fits to the 
$M_\bullet-O$ correlation which treat $M_\bullet$ and $O$ 
symmetrically, such as bisector or orthogonal fits.  Thus, although 
it is commonly used, the $M_\bullet-\sigma$ relation reported by 
Tremaine et al. (2002) is {\em not} the appropriate choice  
for this problem.  

Therefore, we have performed our own fits to the relations we 
require.  The fitting procedure we use is described in the Appendix, 
as are the results of fits to the H\"aring \& Rix (2004) compilation. 

\begin{figure}
 \includegraphics[width=\columnwidth]{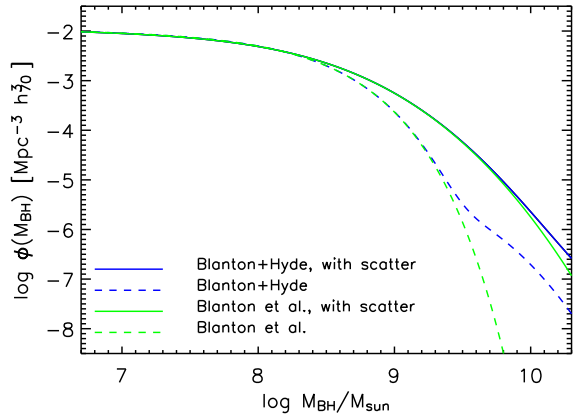}
 \caption{Distribution of black hole masses predicted by combining 
          the $M_\bullet-L$ relation (equation~A\ref{mbhLum}) with the 
          SDSS luminosity function 
          and ignoring (dashed) or including (solid) the effect 
          of 0.33~dex scatter around the mean relation. 
          In each case, bottom curve uses the SDSS luminosity function 
          of Blanton et al. (2003), and the top curve uses Blanton et al. 
          augmented with the BCG luminosities of Hyde et al. (2007).  }
 \label{lbulge2bh}
\end{figure}

\subsection{Effect of scatter in the $M_\bullet-L$ relation}
To estimate $\phi(M_\bullet)$ we need both $p(\log M_\bullet|L)$ 
and the distribution of $L$.  
We use the $r$-band SDSS luminosity function (Blanton et al. 2003) 
as our basic function, augmented so that it includes a better 
estimate of the light from the most luminous galaxies.

Briefly, the SDSS photometric pipeline tends to underestimate 
the luminosities of bright galaxies in crowded fields and of nearby
bright galaxies by more than $0.5$ mag (Bernardi et al. 2007a; 
Lauer et al. 2007; Hyde et al. 2007). The magnitudes of the main 
galaxy sample are biased low by $\sim 0.1$~mag (see Bernardi 2007 
for a discussion of the systematics in the magnitudes and velocity 
dispersion in the SDSS database and comparisons with the 
Bernardi et al. 2003a sample).
Since these bright galaxies are likely to be massive galaxies, they 
are likely to host massive black holes, so it is important to 
correct for this bias.  However, doing so is complicated by the 
fact that the light profiles of these objects are not standard.  
Hyde et al. (2007) believe that the light profiles are the sum of two 
components (a galaxy plus inter-cluster light), and only assign 
the light from the inner component to the object.  
(Assigning all of the integrated surface brightness to the galaxy 
makes the discrepancies described below even larger.)  

The effect of adding these objects to the luminosity function, 
and then transforming to a distribution of black hole masses using 
equation~(A\ref{mbhLum}) is shown by the dashed lines in 
Figure~\ref{lbulge2bh}. 
The effect at the luminous end is dramatic: 
Blanton + Hyde exceeds Blanton alone by many orders of magnitude.

These estimates of black hole abundances ignore the effects of 
intrinsic scatter in the $M_\bullet-L$ relation. The solid curves 
in Figure~\ref{lbulge2bh} show the result of transforming to a 
distribution of black hole masses using equation~(A\ref{mbhLum}) 
and accounting for scatter of 0.33~dex using equation~(\ref{scat}).  
Including the scatter increases the expected $\phi(M_\bullet)$ 
noticably at $M_\bullet > 10^{8.5}M_\odot$; by 
$M_\bullet > 10^{9.5}M_\odot$ ignoring the scatter results 
in an underestimate of more than an order of magnitude.  
In fact, Blanton + scatter exceeds Blanton + Hyde at almost all 
$M_\bullet$.  In this respect, accounting for scatter is more 
important than is getting details of the light profile correct.  

\begin{figure}
 \includegraphics[width=\columnwidth]{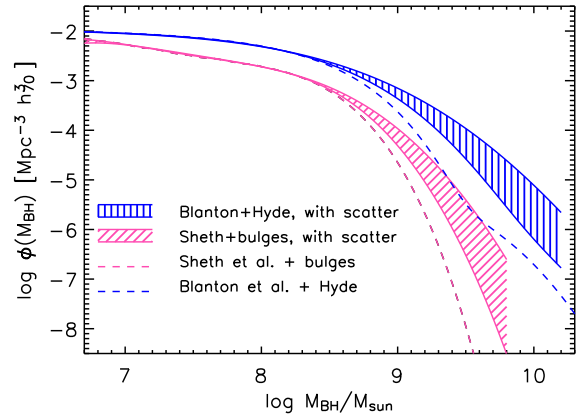}
 \caption{Luminosity and velocity dispersion-based predictions 
          for the distribution of black hole masses.  
          Curves labeled Sheth et al. were obtained by combining 
          the $\langle M_\bullet|\sigma\rangle$ relation of 
          equation~(A\ref{mbhSigma}) with the observed distribution 
          of velocity dispersions (from Sheth et al. 2003).  
          Curves labeled Blanton+Hyde were obtained by combining 
          the $\langle M_\bullet|L\rangle$ relation of 
          equation~(A\ref{mbhLum}) with the observed distribution 
          of luminosity from Blanton et al. (2003) and Hyde et al. (2007).  
          The dashed curves assume there is no intrinsic scatter 
          around the $\langle M_\bullet|{\rm observable}\rangle$ 
          relations, whereas the hashed regions are bounded by 
          curves in which the intrinsic scatter around the 
          relation was assumed to be $0.22\pm0.06$~dex for 
          $\langle M_\bullet|\sigma\rangle$, and $0.33\pm0.08$~dex
          for $\langle M_\bullet|L\rangle$ (see Appendix).  }
 \label{scat2bh}
\end{figure}

\subsection{Abundances from the correlation with $\sigma$}
Figure~\ref{scat2bh} shows the results of repeating this 
analysis, but now with $\langle \log M_\bullet|\log\sigma\rangle$
and the distribution of velocity dispersions reported by 
Sheth et al. (2003).  (A word on this choice is necessary, since 
Bernardi et al. 2006 note that there may be more systems in 
the SDSS with $\sigma\ge 400$~kms$^{-1}$ than the Sheth et al. 
fitting formula yields.  However, HST imaging shows that most of 
the abnormally large $\sigma$ objects in Bernardi et al. (2006) 
are objects in superposition; the shape of the Sheth et al. 
velocity function does {\em not} need to be augmented by more 
systems at $\sigma\ge 400$~kms$^{-1}$).  
For ease of comparison with the luminosity function results shown 
in the previous subsection, we have used $d\phi(\sigma)/d\sigma$ 
shown in the final figure of Sheth et al.---this adds an estimate 
of the contribution of spiral bulges to the measured distribution 
of early-type galaxy velocity dispersions.  Note that this makes 
essentially no difference at the massive end.  

\begin{figure}
 \includegraphics[width=\columnwidth]{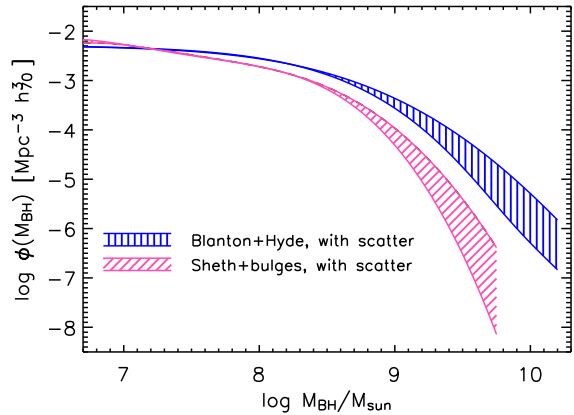}
 \caption{Accounting for the difference between bulge and total 
          luminosity brings the $L$-based estimate of black hole 
          abundances into better agreement with that based on 
          $\sigma$, although the differences at 
          $M_\bullet >10^9M_\odot$ remain.}
 \label{modelBulges}
\end{figure}

\begin{figure*}
 \centering
\includegraphics[width=1.9\columnwidth]{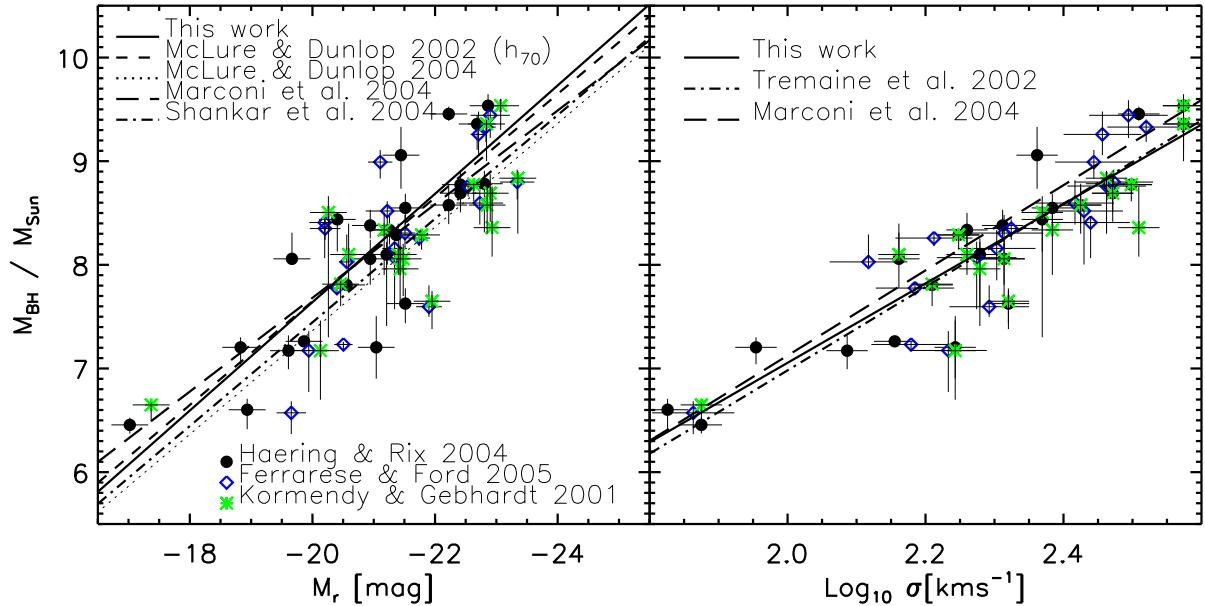}
 \caption{Left: Correlation between $M_\bullet$ and bulge luminosity.  
          Symbols show measurements from a variety of data sets, 
          solid line shows the fit reported in equation~(A\ref{mbhLum}),
          and dashed line shows the fit from McLure \& Dunlop (2002) 
          once the difference in Hubble constant has been 
          accounted for.  Dotted, dot-dashed and long-dashed lines show the 
          fits used by McLure \& Dunlop (2004), Shankar et al. (2004) and
          Marconi et al. (2004), respectively.  
          In all cases, the fits and data have been shifted to the $r$ band
          (using $B-r=1.25$, $V-r=0.34$, $r-R=0.27$, $r-I=1.07$ and $r-K=2.7$).
          Right: Correlation between $M_\bullet$ and velocity dispersion.
          The solid line shows the fit reported in 
          equation~(A\ref{mbhSigma}). Dot-dashed and long-dashed lines 
          show given in Tremaine et al. (2002) and Marconi et al. (2004),
          respectively.}
 \label{MbhLV}
\end{figure*}

The lowest dashed line in the figure shows the expected abundance 
of supermassive black holes if one ignores the intrinsic scatter 
in the $\langle \log M_\bullet|\log\sigma\rangle$ relation, and 
the lower hashed region shows the predicted range if this scatter 
is between 0.16 and 0.28~dex (i.e. $0.22\pm 0.06$~dex, see Appendix).  
The scatter clearly increases the 
expected numbers of massive black holes significantly.  
To appreciate the magnitude of the effect, 
the upper set of curves show the expected abundances based on 
Blanton + Hyde combined with the $\langle M_\bullet|L_r\rangle$ 
relation of equation~(A\ref{mbhLum}) without scatter (upper dashed curve) 
and with scatter (upper hashed region) between 0.25 to 0.41~dex 
(i.e. $0.33\pm 0.08$~dex, see Appendix).  
Notice that the $\sigma$-based prediction when scatter is included 
is similar to the $L$-based prediction when scatter is ignored.

There is a small inconsistency here which we have investigated 
but which does not affect our main conclusion.  Namely, $L$ in the 
$M_\bullet-L$ relations reported earlier refers to the bulge 
luminosity.  Whereas the bulge accounts for all the luminosity at 
large $L$, it accounts for a decreasing fraction at lower $L$.  
We have found that a crude model which sets $L_{\rm bulge} = f(L)\,L$, 
with $f(L) = (L/L_*)/(1 + L/L_*)$ yields bulge luminosity densities 
which are 40\% of the total luminosity density in the $g$ and $r$-bands, 
in good agreement with current estimates.  Figure~\ref{modelBulges} 
shows the result of incorporating this model for $f(L)$ into 
our estimates of $\phi(M_\bullet)$.   Doing so brings the 
$L$- and $\sigma$-based estimates into good agreement at 
$M_\bullet < 10^{7.5}M_\odot$.  However, since $f(L)\to 1$ at 
large $L$, the large differences at $M_\bullet > 10^9\,M_\odot$ remain.  

\section{Problems and inconsistencies}
The smaller intrinsic scatter around 
$\langle M_\bullet|\sigma\rangle$ as compared to 
$\langle M_\bullet|L\rangle$ (equations~A\ref{mbhSigma} 
and~A\ref{mbhLum}) suggests that $r_{MV}>r_{ML}$ (equation~\ref{var}), 
so $p_V(M_\bullet)$ should predict more massive black holes than 
$p_L(M_\bullet)$.  
Figure~\ref{scat2bh} shows the opposite trend:   
the estimate based on the Blanton et al. (2003) luminosity function 
is well in excess of that based on the Sheth et al. (2003) velocity 
dispersion function.  This is true even before adjusting the 
Blanton et al. function upwards at large $L$ to account for BCGs.  
This indicates that something has gone wrong with the logic of 
the previous section.  

Furthermore, the analysis of the previous section suggested that, 
once scatter has been accounted for, both $L$- and $\sigma$-based 
methods should give the same prediction.  Figures~\ref{scat2bh} 
and~\ref{modelBulges} show that the luminosity based predictions 
are still much larger than those based on velocity dispersion.  
In this respect, our findings differ markedly from those of 
McLure \& Dunlop (2004), Shankar et al. (2004) and Marconi et al. (2004) 
who reported that, once scatter had been included, the two estimates 
agree.  As we discuss below, this is because they made different 
choices for the shape and scatter of the $M_\bullet$-observable 
correlations.  Whereas McLure \& Dunlop, and Shankar et al. have 
approximately the same slope as equation~(A\ref{mbhLum}), they are 
shifted to smaller zero-points.  Marconi et al. have a shallower 
slope for $M_\bullet-L$, the zero-point of their $M_\bullet-\sigma$ 
is larger, and they make a different choice for the scatter.  

\subsection{Comparison with previous work}
The left hand panel of Figure~\ref{MbhLV} compares various 
determinations of $\langle M_\bullet|L\rangle$.  (In this figure 
we also show data from Kormendy \& Gebhardt (2001) and 
Ferrarese \& Ford (2005)---although we only use their measurements 
of the objects which are in common with H\"aring \& Rix (see 
Appendix~A in Bernardi et al. 2007b for a decription of
how the black hole sample was compiled).   
We use these other measurements primarily to demonstrate the 
uncertainties on the measurements---all the fits we show and 
use come {\em only} from the H\"aring \& Rix data.)

Both the McLure \& Dunlop (2004) and 
Shankar et al. (2004) results are based on the determination 
of $M_\bullet-L$ by McLure \& Dunlop (2002):  
 $\log M_\bullet = -0.5M_{\rm R} - 2.91$.  
McLure \& Dunlop (2002) say that this determination assumes 
$H_0 = 50$~km~s$^{-1}$~Mpc$^{-1}$.  
Shankar et al. (2004) say that the result of shifting this 
relation to $H_0 = 70$~km~s$^{-1}$~Mpc$^{-1}$ is to change 
the zero point from $-2.91$ to $-2.69$.  This results from 
rescaling {\em both} the luminosities and the black hole masses:  
the net shift is $1.25\log(70/50)^2 - \log(70/50) = 0.22$, 
the first term coming from shifting the luminosities, and the 
second from the masses.  
Note that this rescaling would be appropriate if {\em both} 
$M_\bullet$ and $L$ in the 2002 paper assumed the same $H_0$, 
but would be inappropriate if not.  

Shankar et al.'s (2004) shift differs slightly from that of 
McLure \& Dunlop (2004) who state that, if 
 $H_0 = 70$~km~s$^{-1}$~Mpc$^{-1}$ and $R-K=2.7$, 
then their fit from 2002 implies 
 $\log M_\bullet = 1.25\, \log L_{\rm K}/L_{\odot {\rm K}} - 5.76$.  
Now, $M_{\odot {\rm K}} = 3.28$, so the right hand side of this 
relation is 
 $-0.5\, (M_{\rm R} - 2.7 - 3.28) - 5.76 = -0.5\, M_{\rm R} - 2.77$.  
This relation predicts $M_\bullet$ that are lower by 0.08~dex than 
does the relation used by Shankar et al. (2004).  
The source of this discrepancy is unclear, but it is a curious 
coincidence that $\log(70/50) = 0.146$ is close to the 
$2.91-2.77$ shift that McLure \& Dunlop require:  this would be 
the shift if $M_\bullet\propto L_{\rm R}$ rather than 
$\propto L_{\rm R}^{1.25}$.

\begin{figure}
 \centering
  \includegraphics[width=\columnwidth]{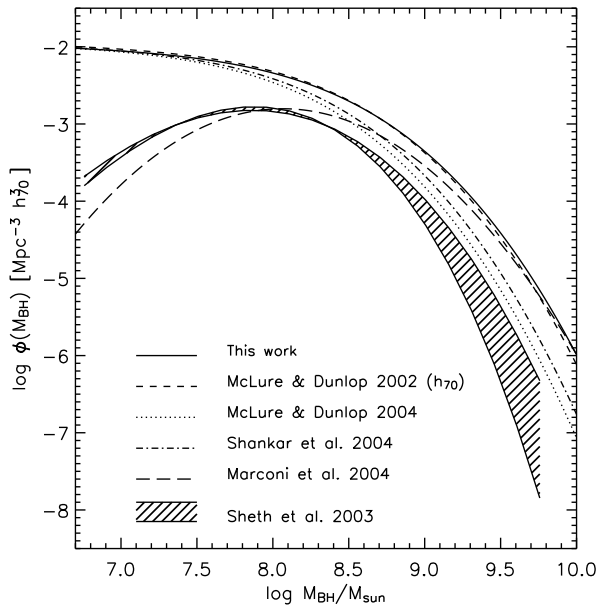}
 \caption{Black hole abundances associated with some of 
          the $M_\bullet-L$ relations shown in the previous figure.
          All curves assume the $r$-band luminosity function from 
          Blanton et al. (2003) except for Marconi et al. (2004) which is
          based on the early-type galaxy sample of Bernardi et al. (2003a).
          Clearly, the fit which produces larger black hole masses for a 
          given luminosity results in the most supermassive black holes. 
          The hashed region labeled Sheth et al. were obtained by combining 
          the $\langle M_\bullet|\sigma\rangle$ relation of 
          equation~(A\ref{mbhSigma}) with the observed distribution 
          of velocity dispersions (from Sheth et al. 2003).
          This region is bounded by curves in which the intrinsic scatter 
          around the relation $\langle M_\bullet|\sigma\rangle$ was assumed 
          to be 0.16 and 0.28~dex. }
 \label{shankar}
\end{figure}

The McLure \& Dunlop (2004) and Shankar et al. (2004) relations 
are shown as the dotted and dot-dashed lines in Figure~\ref{MbhLV}.  
Both have been shifted from $R$ to $r$ using $r-R = 0.27$, and both 
lie below the H\"aring \& Rix data.  
To study why, we returned to the issue of whether or not both 
$L$ {\em and} $M_\bullet$ should have been rescaled when $H_0$ 
was changed.  McLure \& Dunlop (2002) report that their 
$M_\bullet-\sigma$ relation is essentially the same as that of 
Tremaine et al. (2002).  Therefore, they must be using the same 
$M_\bullet$ values as Tremaine et al.  However, the Tremaine et al. 
analysis actually assumes $H_0 = 80$~km~s$^{-1}$~Mpc$^{-1}$ rather 
than 50~km~s$^{-1}$~Mpc$^{-1}$.  To illustrate the effect this has, 
suppose we rescale the $M_\bullet$ values in their 2002 fit by 
(70/80), and the $R$-band luminosities by $(70/50)^2$.  
This would make their relation 
 $\log M_\bullet = -0.5M_{\rm R} - 2.91 + 2.5\log(70/50) - \log(70/80) 
   = -0.5M_{\rm R} - 2.49$; this is a shift in zero point of 0.42.  
Shifting this from $R$ to $r$ using $r-R=0.27$ as before yields 
the short dashed line.  It is in substantially better agreement 
with the H\"aring \& Rix data than is the dotted line.

The relations used by McLure \& Dunlop and Shankar et al. clearly 
produce smaller black holes for a given luminosity.  The most 
important effect of this is to decrease the $L-$based estimate 
of the number of objects with $M_\bullet > 10^9M_\odot$.  This 
is shown in Figure~\ref{shankar}.  
The dotted, dashed and dot-dashed curves show the result of inserting 
the McLure \& Dunlop (2002-2004) and Shankar et al. (2004) based 
$\langle M_\bullet|L\rangle$ relations 
in equation~(\ref{scat}), respectively, 
when the scatter is assumed to be 0.33~dex.  
The solid line shows $\phi(M_\bullet)$ for our fit, and the 
hashed region shows the $\sigma-$based abundances.  Clearly, the 
$\langle M_\bullet|L\rangle$ relation with the smallest zero-point, 
that of McLure \& Dunlop (2004), produces the fewest massive black 
holes.  McLure \& Dunlop are able to account for the small difference 
which remains between the $L$ and $\sigma$-based estimate by 
assigning a larger scatter to the 
$M_\bullet-\sigma$ relation, 0.3~dex, rather than the 
0.22~dex which we used to produce Figure~\ref{shankar}.  
However, the left hand panel in Figure~\ref{MbhLV} suggests that the 
lower zero-point is unacceptably low, and 0.3~dex is larger than all 
recent estimates of the scatter around $\langle M_\bullet|\sigma\rangle$.  

Marconi et al. (2004) also found consistency between the two 
estimates.  They believe that this is because the scatter 
in the $M_\bullet-L$ and $M_\bullet-\sigma$ relations are similar 
(they believe both are about 0.3~dex).  We believe that it is their 
choice of relations combined with the scatter around the relations 
which is the cause of the agreement (the analysis of the previous 
section shows clearly that having equal scatter in the two relations 
is neither sufficient nor necessary).  
To illustrate, the long dashed line in Figure~\ref{shankar} shows the 
$M_\bullet$ distribution computed by Marconi et al. (2004) from the 
$L$-based approach (at lower $M_\bullet$ it differs from the other 
works mainly because Marconi et al. used the early-type galaxy 
luminosity function from Bernardi et al. 2003b instead of the 
luminosity function of all types from Blanton et al. 2003). 
Their $\sigma$-based approach gives a similar curve provided one 
uses their $M_\bullet-\sigma$ relation, shown as the long dashed 
line in the right panel of Figure~\ref{MbhLV}, with intrinsic 
scatter of $0.3$~dex.
The hashed region labeled Sheth et al. was obtained by combining the 
$\langle M_\bullet|\sigma\rangle$ relation of 
equation~(A\ref{mbhSigma}) with the observed distribution of velocity 
dispersions (from Sheth et al. 2003). This region is bounded by curves 
in which the intrinsic scatter around the relation 
$\langle M_\bullet|\sigma\rangle$ was assumed to be 0.16 and 0.28~dex
(note that the larger limit is similar to the value used by
Marconi et al., i.e. 0.3~dex).
However, in this case the $L$ (long dashed
line) and $\sigma$ (upper bound of hashed region) based estimates are
different. Marconi et al. found consistency between the two estimates
because the $M_\bullet-\sigma$ relation they used is steeper and shifted 
to larger $M_\bullet$ values than our relation (see right hand panel of 
Figure~\ref{MbhLV}), so their $\sigma$s produce larger $M_\bullet$s.

Although Marconi et al. (2004), McLure \& Dunlop (2004), and 
Shankar et al. (2004) were able to obtain $L$-based estimates of 
$\phi(M_\bullet)$ which were in good agreement with those based on 
$\sigma$, the analysis above suggests that this was largely due to 
a fortuitous inconsistency resulting from how one rescales $M_\bullet$ 
and $L$ when changing the Hubble constant.  
However, in the next subsection we discuss why, if the Hubble-constant 
related scalings are all done self-consistently, then the $\sigma-$ 
and $L-$based estimates should {\em not} have given the same answer!

\subsection{The $\sigma-L$ relation}
\begin{figure}
 \includegraphics[width=\columnwidth]{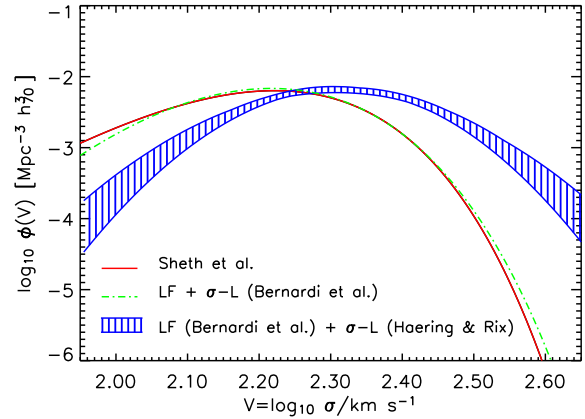}
 \caption{Observed and predicted distribution of $\sigma$;  
          solid curve shows the velocity function reported by 
          Sheth et al. (2003); 
          dot-dashed line shows the result of starting from the 
          luminosity function of Bernardi et al. (2003b), and using 
          the SDSS $\langle\sigma|L\rangle$ relation and its scatter 
          to infer $\phi(\sigma)$, whereas hashed region uses the 
          $\langle\sigma|L\rangle$ relation and scatter from the 
          black hole sample of H\"aring \& Rix (2004) instead.  
          The $\langle\sigma|L\rangle$ relation for the black hole 
          sample is rather uncertain, since it is derived from only 
          $\sim 30$ objects: the hashed region shows the range of 
          predicted $\phi(\sigma)$ associated with allowing the slope 
          and scatter of the $\sigma-L$ relation to vary within one 
          standard deviation of their rms values.  }
 \label{wrongV}
\end{figure}

Why do our $L-$ and $\sigma-$based estimates give different answers?  
If we transform the SDSS-B07 luminosity distribution into one for 
$\sigma$ using equations~(A\ref{sigLum}) and~(\ref{scat}), and then 
to a distribution of $M_\bullet$ using equations~(A\ref{mbhSigma}) 
and~(\ref{scat}), then this gives the same answer as transforming 
SDSS-B07 luminosity into $M_\bullet$ directly using equations~(A\ref{mbhLum}) 
and~(\ref{scat}).  This is exactly as expected from the toy model 
described in the previous section.  However the intermediate step 
provides a predicted velocity function which disagrees with the 
SDSS-B07 one (from Sheth et al. 2003).  
Figure~\ref{wrongV} shows this explicitly; the hashed region shows 
the result of starting with the SDSS $\phi(L)$ and using the black 
hole $\langle\sigma|L\rangle$ relation and scatter 
(equation~(A\ref{sigLum})) to infer $\phi(\sigma)$.  
The range of values comes from including the uncertainty in the 
slope and scatter of $\langle\sigma|L\rangle$.  The disagreement 
with the actual measured $\phi(\sigma)$ distribution (solid curve) 
strongly suggests that the $\sigma-L$ relation in the black 
hole samples is not the same as in the SDSS-B07 sample, and that 
this is the source of the discrepancy between the $L-$ and 
$\sigma-$based estimates.  

In SDSS-B07,
\begin{equation}
 \Bigl\langle \log\sigma|M_r\Bigr\rangle_{\rm SDSS-B07} 
        = 2.287 - {0.255\over 2.5}\,(M_{r}+22),
 \label{sdssVL}
\end{equation}
with an error in the slope and zero-point of $0.009$ and $0.005$, 
respectively, whereas it is 
\begin{equation}
 \Bigl\langle \log \sigma|M_r\Bigr\rangle 
        = 2.42 - {0.34\over 2.5}\,(M_{r}+22)
 \label{bhVL}
\end{equation}
in the H\"aring \& Rix sample (equation~A\ref{sigLum}).  
The errors in the slope and zero-point are $0.02$ and $0.01$, 
respectively.
Note that this slope of $-0.34/2.5=-0.14$ is rather different from the 
canonical value of $-0.10$:  
At a given luminosity, the black hole samples have $\log\sigma$ 
larger by about 0.08~dex than the SDSS-B07---observational errors are 
typically only about $0.03$~dex.  

Yu \& Tremaine (2002) also considered the possibility that the 
$\sigma-L$ relation was the cause of the discrepancy, and suggested 
that perhaps there are systematic differences between SDSS-B07 velocity 
dispersions and those derived from more local samples.  A direct 
test of this possibility is difficult because, of the $\sim 30$ 
objects in the H\"aring \& Rix compilation, only about ten have SDSS 
imaging, and only NCG 4261 has an SDSS spectrum as well.  
For the objects in common, the SDSS apparent magnitudes are about 
0.5~mags fainter than those used in the black hole analyses, but 
this is almost certainly due to the sky subtraction problems for 
bright objects to which we refered earlier (Hyde et al. 2007).  
In any case, correcting for this will increase the SDSS luminosities, 
further exacerbating the discrepancy in the $\sigma-L$ relation.  

\begin{figure}
 \includegraphics[width=\columnwidth]{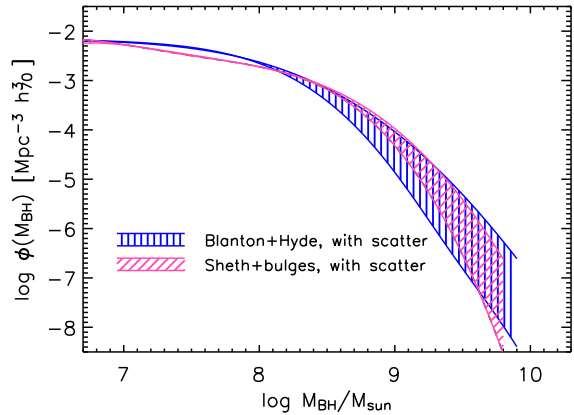}
 \caption{Predicted abundances if the luminosities of the black 
          hole hosts are modified so that they define a 
          $\langle\sigma|L\rangle$ relation which has the same 
          slope and zero-point as the SDSS-B07 relation.  
          This rescaling modifies the 
          $\langle M_\bullet|L\rangle$ relation, but leaves 
          the $\langle M_\bullet|\sigma\rangle$ relation unchanged.  
          As a result the curve labeled `Sheth + bulges' is the 
          same as before, but `Blanton + Hyde' now produces many 
          fewer massive objects.  }
 \label{wrongL}
\end{figure}

A detailed discussion of the $\sigma-L$ relation computed from 
different samples, analysis of systematic biases which affect the 
samples, and the effect of correcting ``naively'' the nearby 
samples for peculiar velocities, is presented in Bernardi (2007). 
Compared to any of these early-type galaxy samples, 
the $\sigma-L$ relation in black hole samples is biased to larger 
$\sigma$ for given $L$, or to smaller $L$ for a given $\sigma$ 
(Bernardi et al. 2007b).  
In view of this discrepancy, whatever the cause, the fact that 
McLure \& Dunlop (2004), Marconi et al. (2004) and Shankar et al. (2004) 
obtained consistent estimates of $\phi(M_\bullet)$ from both $L$ and 
$\sigma$ is remarkable indeed.  

Figure~\ref{wrongL} shows the result of assuming that the velocity 
dispersion estimates in the black hole sample are reliable, but 
the distances, and so the luminosities, are not.  
It was constructed by rescaling all the bulge luminosities of 
the black hole hosts so that they define a relation with the 
same slope as in equation~(\ref{sdssVL}), though 
with different scatter.  To do so, we added $-0.978 + 0.25\,(Mr+22)$ 
to each of the absolute magnitudes in the black hole sample, 
as suggested by the difference between equations~(\ref{sdssVL}) 
and~(\ref{bhVL}).

These rescaled luminosities were used to estimate a new 
$\langle M_\bullet|L_r\rangle$ relation, 
which was then inserted in equation~(\ref{scat}) to predict 
black hole abundances from the luminosity function.  The 
resulting abundances are considerably lower, because the 
rescaled luminosities define a considerably shallower 
$\langle M_\bullet|L_r\rangle$ relation, meaning that considerably 
larger $L$ is required to reach $M_\bullet>10^9M_\odot$.  
While this rescaling is probably unrealistic, we have included 
the result to illustrate the importance of the $\sigma-L$ 
relation when comparisons of the $L-$ and $\sigma-$based 
estimates of $\phi(M_\bullet)$ are made.  A more careful 
accounting of the role of selection effects is presented in
Bernardi et al. (2007b).  

\section{Discussion}
It is common to estimate the abundance of supermassive black holes 
by combining observed correlations between $M_\bullet$ and bulge 
luminosity or velocity dispersion, calibrated from relatively small 
samples, with luminosity or velocity dispersion functions determined 
from larger samples.  However, the $\langle M_\bullet|\sigma\rangle$ 
and $\langle M_\bullet|L\rangle$ relations have intrinsic scatter of 
about 0.22 and 0.33~dex (Appendix).  
Accounting for this results in considerably increased estimates 
of the abundance of black holes with $M_\bullet \ge 10^9\,M_\odot$, 
compared to naive estimates which ignore this scatter.  
Doing so is at least as important as correcting the luminosity 
function for the fact that the most luminous galaxies have 
non-standard light profiles (Figure~\ref{lbulge2bh}).  
Once this scatter has been accounted for, the $\sigma$-based 
estimates of $\phi(M_\bullet)$ are in reasonably good agreement 
with models, such as that of Hopkins et al. (2006), which relate 
previous QSO and AGN activity to the local black hole mass function.  
The luminosity-based estimates, on the other hand, are substantially 
in excess of this model.  

These results follow from using a single power-law to parametrize 
the $M_\bullet-\sigma$ and $M_\bullet-L$ relations.  While this may 
be too simplistic, this parametrization is not the primary reason 
why the $L-$ and $\sigma-$based approaches yield different 
predictions for black hole abundances.  
The main cause of the discrepancy is that the $\sigma-L$ correlation 
in black hole samples is different from that in the samples from which 
the luminosity and velocity functions are drawn:  the black hole 
samples have larger $\sigma$ for a given $L$ compared to the ENEAR 
or SDSS-B07 samples or have smaller $L$ for a given $\sigma$ 
(Bernardi et al. 2007b).  
If this is a physical effect, then it compromises the fundamental 
assumption of black hole demographic studies---that all galaxies 
host black holes.  If, on the other hand, it is a selection effect, 
then the $M_\bullet-\sigma$ and $M_\bullet-L$ relations currently 
in the literature are biased compared to the true relations, 
making current estimates of black hole abundances unreliable.  
If black hole masses correlate with bulge luminosity only because 
of the $M_\bullet-\sigma$ and $\sigma-L$ relations, then the bias 
in the $\sigma-L$ relation is not important only if one is using 
$\phi(\sigma)$ to infer black hole abundances:  the 
$\phi(L)$-based estimate may be strongly affected.

Identifying the source of the bias is complicated.  
Residuals from the size-luminosity relation are anti-correlated 
with residuals from the $\sigma-L$ relation, as might be expected 
from the virial theorem (Bernardi et al. 2003b).  
If the stellar kinematics method of measuring black hole masses 
favors objects with high surface brightnesses, then one might 
expect smaller sizes and larger $\sigma$ at constant $L$:  this 
would produce a bias in the sense we see.  
On the other hand, it might be more difficult to measure the 
influence of the black hole on stellar kinematics if the stellar 
velocity dispersion is already abnormally large---this would
produce a bias in the opposite sense.  Whether either or these 
effects has played a role in the selection of black samples is 
an open question.  
See Bernardi et al. (2007b) for further study along these lines.

\acknowledgements

This work is partially supported by NASA grant 
LTSA-NNG06GC19G, and by grants 10199 and 10488 from the Space 
Telescope Science Institute, which is operated by AURA, Inc., 
under NASA contract NAS 5-26555.

Funding for the SDSS and SDSS-II
has been provided by the Alfred P. Sloan Foundation, 
the Participating Institutions, the NSF, the US DOE, NASA, 
the Japanese Monbukagakusho, the Max Planck Society 
and the Higher Education Funding Council for England.  
The SDSS website is http://www.sdss.org/.

\appendix
This Appendix describes our procedure for estimating the slope 
and scatter associated with $\langle M_\bullet|O\rangle$.  
Let $y = \log M_\bullet - \langle \log M_\bullet\rangle$, and 
$x = O - \langle O\rangle$, and let $\sigma_x$, $\sigma_y$ and 
$r_{xy}$ denote the true intrinsic rms values of $x$, $y$, and the 
cross-correlation coefficient.  Finally, let $\epsilon_x$ and 
$\epsilon_y$ denote the typical measurement errors in determining 
$x$ and $y$.  In practice, the estimated error may vary from object 
to object; in using a single representative value, our analysis 
below ignores this additional information.  
Minimizing
\begin{equation}
 \chi^2 \equiv \sum_{i=1}^N (y_i - a x_i - b)^2/N
\end{equation}
with respect to $a$ and $b$ yields 
\begin{equation}
 a_{\rm min} = {\sum_i x_iy_i\over \sum_i x_i^2} 
     = {\sigma_x\sigma_y r_{xy}\over \sigma_x^2 + \epsilon_x^2} 
     = {\sigma_x\sigma_y r_{xy}\over \sigma_x^2} 
       \left(1 + {\epsilon_x^2\over\sigma_x^2} \right)^{-1}
     = a_{y|x}\,
       \left(1 + {\epsilon_x^2\over\sigma_x^2} \right)^{-1}
\end{equation}
and, because both $x$ and $y$ have zero mean, $b_{\rm min}=0$.  
Comparison with equation~(\ref{mean}) shows that $a_{\rm min}$ 
differs from the true slope $a_{y|x}$ because of the measurement 
errors $\epsilon_x^2$.  (We have assumed uncorrelated measurement 
errors in $x$ and $y$.  Hence, these errors affect the mean of 
$x^2$, and of $y^2$, but not the correlation between $x$ and $y$.)
The scatter around this relation is 
\begin{equation}
 \chi^2_{\rm min} = \sigma_x^2\, (1-r_{xy}^2) + \epsilon_y^2  
  + a_{\rm min}^2\, \epsilon_x^2 
   \left(1 + {\epsilon_x^2\over\sigma_x^2}\right).
\end{equation}
The uncertainty on the value of this minimum is 
$\chi^2_{\rm min} \sqrt{2(N-1)/N^2}$:  for $N=24$ the estimated 
scatter is uncertain by about $\sqrt{46}/24\sim$ thirty percent.  

Comparison with equation~(\ref{var}) shows that the first term 
in the expression above represents the intrinsic scatter around 
the true relation, and the other terms are a consequence of the 
measurement errors.  
Hence, the intrinsic slope and scatter which we report in the 
main text are 
\begin{equation}
 a_{\rm y|x} = a_{\rm min}\, 
                \left(1 + {\epsilon_x^2\over\sigma_x^2} \right)\qquad
 {\rm and}\qquad 
 \sigma^2_{\rm y|x} = \chi^2_{\rm min} - \epsilon_y^2  
  - a_{\rm min}^2\, \epsilon_x^2 
    \left(1 + {\epsilon_x^2\over\sigma_x^2}\right).
\end{equation}
Notice that $a_{y|x}$ can be determined well even if $\epsilon_y$ 
is large; of course large uncertainties in $y$ do affect the 
scatter around the mean relation.  
There will be trouble only if $\sigma_x\ll \epsilon_x$; 
in this case, the large measurement errors in $x$ have largely 
erased the correlation between $x$ and $y$, so a small measured 
slope requires a large correction factor to restore it to the 
true value.  

We have applied this procedure to the dataset of H\"aring \& Rix (2004), 
who provide estimates of $M_\bullet$, $\sigma$, 
$M_{\rm bulge}$ and $L_r$ and the fraction of this luminosity which 
is from the bulge (Appendix~A in Bernardi et al. 2007b describes exactly 
how the black hole sample was compiled and the conversion from B, V, R and 
I-band luminosities to SDSS $r-$band. Both luminosities and the black hole 
masses were scaled to $H_0 = 70$~km~s$^{-1}$~Mpc$^{-1}$). 
When doing so we will deal almost exclusively with logarithmic quantities; 
when taking the logarithm, $M_\bullet$ 
is in units of $M_\odot$, $\sigma$ is in kms$^{-1}$, and the associated 
measurement errors are   
$\epsilon_{\log M_\bullet} \sim 0.2$~dex, 
$\epsilon_{\log \sigma} \sim 0.03$~dex, and 
$\epsilon_{\log M_{\rm bulge}} \sim 0.18$~dex.  
The scatter around the correlations we report are estimates of the 
intrinsic scatter. The uncertainties in the slope, zero-point, and scatter
of the following relations were computed by bootstrap resampling.
Application of the procedure outlined above yields 
\begin{equation}
 \Bigl\langle \log M_{\bullet}|\log \sigma\Bigr\rangle 
  = (8.21 \pm 0.06) + 
 (3.83 \pm 0.21) \,\log\left({\sigma\over 200\,{\rm kms}^{-1}}\right)
 \label{mbhSigma}
\end{equation}
with intrinsic scatter of $0.22\pm 0.06$~dex, and 
\begin{equation}
 \Bigl\langle \log M_{\bullet}|\log M_{\rm bulge}\Bigr\rangle = 
    (8.31 \pm 0.10) +
    (1.06 \pm 0.12)\,\log \left({M_{\rm bulge}\over 10^{11}M_\odot}\right)
 \label{mbhBulge}
\end{equation}
with rms scatter $0.33\pm 0.08$~dex.  
Bulge mass and luminosity are tightly correlated (H\"aring \& Rix 2004):
\begin{equation}
 \Bigl\langle \log M_{\rm bulge}|M_r\Bigr\rangle = 11.35 - 0.492\,(M_r + 22)
\end{equation}
with negligible scatter, so inserting this fit in the previous one 
yields 
\begin{equation}
 \Bigl\langle \log M_{\bullet}|M_r\Bigr\rangle 
  = 8.69 - {1.31\over 2.5}\,(M_{r}+22)
\end{equation}
with scatter of $0.33$~dex.  
As a check, we have also fit for the correlation between $M_\bullet$ 
and $M_r$ directly, finding 
\begin{equation}
 \Bigl\langle \log M_\bullet|M_r\Bigr\rangle 
        = (8.68\pm 0.10) - {(1.30\pm 0.15)\over 2.5}\,(M_{r}+22)
 \label{mbhLum}
\end{equation}
with scatter of $0.34\pm 0.09$~dex.  
The main text also considered the correlation between $L$ and 
$\sigma$ in this data set.  It is 
\begin{equation}
 \Bigl\langle \log \sigma|M_r\Bigr\rangle 
        = (2.42\pm 0.01) - {(0.34\pm 0.02)\over 2.5}\,(M_{r}+22)
 \label{sigLum}
\end{equation}
with scatter of $0.04\pm 0.01$~dex.  
Note that this slope is rather different from the 
canonical value of $-0.10$.

\end{document}